\documentstyle[aps,epsf,floats]{revtex}

\ifx\epsffile\undefined
\def\insertfig#1{}
\else
\def\insertfig#1{{\baselineskip=4pt
\centerline{\epsfxsize=\hsize\epsffile{#1}}}}\fi

\begin{document}

\twocolumn[       
{\tighten
\preprint{\vbox{
\hbox{}
\hbox{}
}}
\draft
\title{\Large\bf Ghost poles in the nucleon propagator in the linear 
$\bbox{\sigma}$ model approach and its role in $\bbox{\pi}{N}$ low-energy 
theorems}
\author{C.A. da Rocha$^{(1)}$\footnotemark, G. Krein$^{(2)}$\footnotemark,
and L. Wilets$^{(1)}$\footnotemark}
\address{$(1)$ Department of Physics, Box 351560, University of Washington, Seattle, 
Washington 98195-1560 \\
$(2)$ Instituto de F\'{\i}sica Te\'orica - Universidade Estadual Paulista - 
Rua Pamplona 145, 01405-900 S\~ao Paulo-SP, Brazil}
\bigskip
\bigskip
\date{January 1997}
\maketitle
\widetext
\vskip-2.25in
\rightline{\fbox{
\hbox{DOE/ER/40427-26-N96}
}}
%
\vskip 1.8in

\begin{abstract}
Complex mass poles, or ghost poles, are present in the Hartree-Fock solution
of the Schwinger-Dyson equation for the nucleon propagator in renormalizable
models with Yukawa-type meson-nucleon couplings, as shown many years
ago by Brown, Puff, and Wilets (BPW). These ghosts 
violate basic theorems of quantum field theory and their origin is related to 
the ultraviolet behavior of the model interactions. Recently, Krein
et.al, proved that the ghosts disappear when 
vertex corrections are included in a self-consistent way, softening the 
interaction sufficiently in the ultraviolet region. In previous 
studies of $\pi N$ scattering using ``dressed'' nucleon propagator and bare 
vertices, did by Nutt and Wilets in the 70's (NW), it was found 
that if these poles are explicitly included, the value of the isospin-even 
amplitude $A^{(+)}$ is satisfied within 20\% at threshold. The absence 
of a theoretical explanation for the ghosts and the lack of chiral 
symmetry in these previous studies 
led us to re-investigate the subject using the approach of the linear
$\sigma$-model and study the interplay of 
low-energy theorems for $\pi N$ scattering and ghost poles. 
For bare interaction vertices we find that ghosts are present in this model as 
well and that the $A^{(+)}$ value is badly described. As a first 
approach to remove these complex poles, 
we dress the vertices with phenomenological form factors and a 
reasonable agreement with experiment is achieved. In order to fix the two 
cutoff parameters, we use the $A^{(+)}$ value for the chiral limit 
($m_\pi \rightarrow 0$) and the experimental value of the isoscalar 
scattering length. Finally, we test our model by 
calculating the phase shifts for the S waves and we find a good agreement at
threshold.
\end{abstract}

}

] 
\narrowtext

\footnotetext{${}^*$Fellow from CNPq Brazilian Agency. Electronic address:
carocha@phys.washington.edu}
\footnotetext{${}^\dagger$Electronic address: gkrein@axp.ift.unesp.br}
\footnotetext{${}^\ddagger$Electronic address: wilets@phys.washington.edu}

\section{INTRODUCTION}
\label{sec:introduction}

Elastic pion-nucleon ($\pi N$) scattering has been studied for more than 
40 years. The body of work until the early 80's was reviewed by 
H\"ohler~\cite{Hol83}. In recent years a renewed interest in $\pi N$ scattering
is being witnessed~\cite{PJ91,GS93,GLM93} in the literature. This
interest is driven mainly by the necessity of having a relativistic description
of the available high energy data, as well as of the data to be generated at
CEBAF. Also, the recognition of chiral symmetry as a fundamental symmetry of 
the strong interactions has motivated a great deal of attention to the role of
this symmetry in the $\pi N$ process. This last point is particularly 
interesting in view of the possibility offered by the $\pi N$ process for 
studying the interface between hadron and quark dynamics. 

In principle, all properties of hadronic processes should be derivable from 
the fundamental theory of the strong interactions, Quantum Chromodynamics 
(QCD). However, the mathematical complexities presented by QCD forbids the 
direct use of this theory for treating strongly interacting processes at low 
energies. Therefore, in practice one is required to use models with effective 
degrees of freedom. In this respect, the use of relativistic quantum field 
models with baryon and meson degrees of freedom for studying low energy 
hadronic processes is common practice. In fact, such models have been used for
a long time for treating nucleon-nucleon (NN) processes, and have been
reasonably successful in the description of the empirical data \cite{NN,MHE}.
It is therefore natural to expect that such hadronic models are adequate for
treating the $\pi N$ processes. On the other hand, it is clear that a hadronic
description must break down for those observables which receive short-distance
contributions, i.e., contributions from distances where quark and gluon degrees
of freedom are directly involved. Whereas for those observables that are 
thought to be insensitive to the short distance physics, such a hadronic
approach should be adequate. However, there is one major difficulty one must
face when using relativistic quantum field models, namely the problem of
renormalization of the ultraviolet divergencies that plague the models. 
Three different renormalization attitudes are usually followed in the
literature:  (i) the ultraviolet divergencies are cutoff by parameters adjusted
phenomenologically, (ii) the model is renormalizable in the sense that it has
a finite number of coupling constants that are fixed by a finite number of
counter terms necessary to eliminate the ultraviolet divergencies, (iii) the
model has an infinite number of coupling constants and the ultraviolet
divergencies are 
absorbed in the redefinition of the coupling constants order by order in the 
expansion in powers of the momenta involved in the process. Attitude (i) is a 
pragmatic one; the cutoff is used to eliminate the short distance physics 
associated with the models, since baryons and mesons are composite particles 
the short distance physics is properly treated once their quark and gluon 
structure is taken into account. The use of renormalizable models, attitude 
(ii), is theoretically preferred since the number of parameters is finite and 
has been proved to be very successful in perturbative calculations of QCD and 
electroweak processes. Attitude (iii) has been pursued recently in the context 
of chiral perturbation theory~\cite{GSS88,JM91,Eck94,Leu99,Wei91}, and has the
potentiality to become useful if higher order calculations, which are required
for intermediate to high energy hadronic processes, can be implemented in
practice.

The present paper starts with attitude (ii). As said above, one advantage
of using renormalizable models is that they are characterized by a finite 
number of coupling constants and therefore contributions from quantum 
fluctuations of the fields can be calculated without introducing additional 
short-distance cutoffs. The insensitivity of low energy observables
to the short distance physics associated with the models is obviously crucial
for the success of such an approach. In this paper we examine the  
insensitivity to short distance physics of a renormalizable hadronic model by
examining the effects of quantum fluctuations for the nucleon propagator in 
low energy $\pi N$ scattering. An important question in the process of 
calculating quantum fluctuations for propagators is the problem of appearance 
of complex poles, or {\em ghost poles}, in the renormalized nucleon and meson 
propagators. The appearance of ghost poles have long been noted in local 
relativistic theory~\cite{long,Wil79}.
The presence of ghost poles in the propagators violates basic theorems of 
local quantum field theory, and the ghosts are 
physically unacceptable because they correspond to eigenstates of the system 
with complex energies and probabilities.  In the case of quantum 
electrodynamics (QED), the complex pole in the photon propagator is known as 
the Landau ghost. The presence of the Landau ghost is not taken as a serious 
drawback of QED since the momentum scale at which it appears is far from 
measurable and, at this scale QED should probably be modified to include other
electroweak effects. However, for hadronic models the ghosts are a problem,
since they appear in the meson and nucleon propagators at a scale as low as 
the nucleon mass. 

\begin{figure}
\vspace{0.5cm}
\epsfxsize=8.0cm
\epsfysize=7.0cm
\centerline{\epsffile{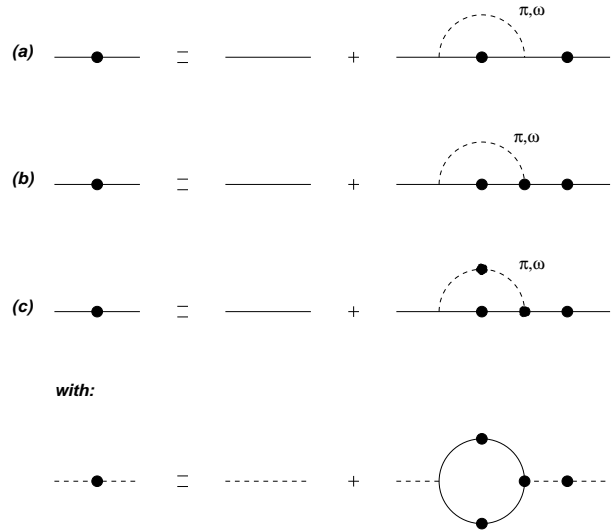}}
\vspace{0.6cm}
\caption{Diagrammatic representation of the Schwinger-Dyson equations:
for the nucleon propagator (a) in the Hartree-Fock approximation and (b) 
with form-factor dressing of the vertices as in Ref.~\protect\cite{Kre+93}, and
(c) for the meson and nucleon propagators as in Ref.~\protect\cite{self}. 
The solid lines refer to nucleons and the dashed lines refer to mesons. A blob
indicates nonperturbative quantities.}
\label{Fig.1}
\end{figure}

Brown, Puff and Wilets (BPW) \cite{BPW} calculated the renormalized nucleon 
propagator in the Hartree-Fock (HF) approximation, which amounts to summing
all planar meson diagrams with one nucleon line as shown schematically in 
Fig.~\ref{Fig.1}(a). The renormalized propagator was well defined and
self-consistent, but contained a pair of ghost poles
located approximately 1 GeV off the real and complex axes. BPW have also shown
that in the HF approximation the ghosts come from the ultraviolet behavior of
the nucleon-meson interaction. This is emphasized by the fact that
asymptotically free models appear to be free of ghost poles~\cite{asymp}.
Recently, Krein, Nielsen, Puff, and Wilets
(KNPW)~\cite{Kre+93} resumed the study of ghost poles in the nucleon
propagator. Following earlier studies by Milana~\cite{JM} and Allendes and 
Serot~\cite{AllSer}, KNPW investigated the effect on the ultraviolet behavior 
of the interaction by the dressing of the nucleon-meson vertices by means of 
the neutral vector meson $\omega$. The vector-meson dressing of the vertices
gives rise to a form factor similar to the Sudakov form factor in QED and has
the effect of softening the ultraviolet behavior 
of the interaction and killing the
ghosts. This vertex dressing is shown schematically in Fig.~\ref{Fig.1}(b). 
In a more recent paper~\cite{self} the coupled SDE's for the nucleon and meson
propagators were solved self-consistently in an approximation that goes beyond
the Hartree-Fock approximation, see Fig.~\ref{Fig.1}(c). The main result
was that the positions in the complex plane of the ghost poles of both nucleon
and meson propagators are rather insensitive to the self-consistency.

The first study of $\pi N$ scattering using the ``dressed'' nucleon 
propagator was done by Nutt and Wilets (NW)~\cite{nutwil} in the 
context of HF approximation. NW used a renormalizable hadronic model with
pseudoscalar pion-nucleon coupling, including the $\omega$ meson, to 
calculate $\pi N$ scattering amplitudes. Using the pole approximation 
for the nucleon propagator (just the first graph of Fig.~\ref{Fig.1}(a) 
after the equal sign), they found the familiar bad result of a zero value 
for the isospin-even amplitude $A^{(+)}$. Besides, the inclusion of just the 
nucleon pole causes a serious problem in that it produces very large S-wave 
$\pi N$ amplitudes which leads to an unphysical behavior in the scattering 
length and in the total cross section near threshold. Including the nucleon 
self-energy term, i.e., including the non-delta function part of spectral 
function, the result becomes $A^{(+)} \simeq -0.17 g_{\pi}^2/M$, where
$g_\pi$ is the pion-nucleon coupling constant and $M$ 
is the nucleon mass. Including the ghost poles, i.e., using the entire 
spectral function of the nucleon propagator, NW found a much better agreement.
The result of NW is $A^{(+)}=0.99 g_{\pi}^2/M$ for the ghosts alone, and
$A^{(+)}=0.82 g_{\pi}^2/M$ for the total contribution, while the Adler
theorem gives $A^{(+)} \simeq  g_{\pi}^2/M$ \cite{Adl65,Wei66}. 
The interpretation of this 
result is that the role of the ghost pole is to cancel the unphysical
behavior of the low-energy observables.
The rationale for the NW result was that the ghost poles
must be included in the calculations for reasons of consistency,
although violating some basic theorems of quantum field theory.

The absence of chiral symmetry in this first study led us to re-investigate 
the problem using a chiral lagrangian, such as the linear sigma model of 
Gell-Mann and Levy~\cite{sigm}, to treat the $\pi N$ scattering. 
Several questions can be raised with respect to the interplay of 
low-energy theorems
and ghost poles. One particularly interesting question is the interplay
of the chiral partner of the pion, the $\sigma$ meson, and the presence of 
ghost poles in the nucleon propagator. This and related questions are
investigated in the next sections. 

At this point we would like to make clear that our studies in this paper
do not intend to describe the details of experimental data. Instead, we
investigate the role of the ultraviolet behavior of the interactions on low
energy observables and hope that the lessons drawn form this study will be of
relevance for future, more sophisticated calculations. As mentioned above, the
problem of constructing a consistent, ghosts free effective field-theory of
mesons and nucleons will certainly be required for higher energies and 
our study is a first step towards such an effort. 

We start Sec.~\ref{sec:review} presenting
the model we use and briefly reviewing the past work on
ghost poles. Next we study the role of chiral symmetry in this approach through
the introduction of a scalar-isoscalar meson in the SDE equations. We begin
Section~\ref{sec:interplay} with a review on $\pi N$ scattering kinematics and 
discuss the role of chiral symmetry in $\pi N$ scattering,
observing the results for the scattering length observable. Next we dress the 
nucleon propagator and re-evaluate the $a^+$ observable in two approaches: 
first, in the spirit of attitude (ii) already described,
we use bare interaction vertices and obtain a nucleon propagator 
self-consistently, which is used to study the low-energy theorems of $\pi N$
interaction. After this, we move to attitude (i) and, using phenomenological 
form factors to dress the vertices, we repeat the same calculations adding a 
evaluation of $\pi N$ phase shifts to test the cutoff parameters of the model.
The conclusions are presented in Section~\ref{sec:conclusions}. 

\section{The model, the nucleon propagator, and ghost poles}
\label{sec:review}
                                            
\subsection{The model}

The Lagrangian density of the linear-$\sigma$ model, augmented by the $\omega$
meson, is given by:
\begin{equation}
{\cal L}={\cal L}_0+\epsilon {\cal L}_1 \;,
\end{equation}
\noindent with
\begin{eqnarray}
{\cal L}_0&=&\bar{\psi}\left[i\gamma^\mu\partial_\mu-g_0\left(\sigma+i\gamma_5
\bbox{\tau}\cdot\bbox{\pi}\right)-g_{0\omega}\gamma_{\mu}\omega^{\mu}\right]
\psi \nonumber \\ [0.3cm]
&+&\frac{1}{2}\left(\partial_{\mu}\sigma\;\partial^{\mu}\sigma
                     +\partial_{\mu}\bbox{\pi}\cdot\partial^{\mu}\bbox{\pi} 
                \right)-\frac{1}{4}F_{\mu\nu}F^{\mu\nu}
\nonumber \\ [0.3cm]
&+&\frac{1}{2}m_\omega^2\omega_{\mu}
\omega^{\mu}
-\frac{\mu_0}{2}\left(\sigma^2+\bbox{\pi}^2\right)-\frac{\lambda^2}{4}
\left(\sigma^2+\bbox{\pi}^2\right)^2 \label{lags} \\ [0.6cm]
{\cal L}_1&=&c\sigma \;,
\end{eqnarray}
where $\psi$ is the nucleon field operator, and $\bbox{\pi}$, $\sigma$ and
$\omega$ are the pseudoscalar-isovector, scalar-isoscalar and
vector-isoscalar meson field operators respectively, $\mu_0$ is the 
degenerated mesons mass,
and $F_{\mu\nu}=\partial_{\mu}\omega_{\nu}-\partial_{\nu}\omega_{\mu}$. The 
${\cal L}_1$ term is the symmetry-breaking term which reproduces PCAC, 
$\epsilon$ being a small parameter. The invariance of ${\cal L}_0$ under chiral 
transformations is described in detail by Lee \cite{Lee72}.

Due to the fact that ${\cal L}_1$ is linear in 
the field $\sigma$, it allows a nonvanishing vacuum expectation
value of the $\sigma$-field. Let $v$ be vacuum expectation value of 
$\sigma$:
\begin{equation}
<\sigma>_0=v\;.
\end{equation}
\noindent
We may define a new field $s$ by the equation
\begin{equation}
\sigma=s+v
\end{equation}
\noindent so that
\begin{equation}
<s>_0=0\;.
\label{Eq.s}
\end{equation}
\noindent
Rewriting the lagrangian in terms of the new fields $s$ we get
\begin{equation}
{\cal L}={\cal L}_a+\epsilon{\cal L}_b 
\end{equation}
\noindent with
\begin{eqnarray}
{\cal L}_a&=&\bar{\psi}\left[i\gamma_\mu\partial^\mu-M-g_0\left(s+i
\gamma_5\bbox{\tau}\cdot\bbox{\pi}\right)\right]\psi
\nonumber \\ [0.3cm]
&+&\frac{1}{2}\left(
\partial_{\mu}\bbox{\pi}\cdot\partial^{\mu}\bbox{\pi}-m_{\pi}^2\bbox{\pi}^2
\right)+\frac{1}{2}\left(\partial_{\mu}s\;\partial^{\mu}s-m_{\sigma}^2
\sigma^2\right)
\nonumber \\ [0.3cm]
&-&\lambda^2vs\left(\sigma^2+\bbox{\pi}^2\right)-
\frac{\lambda}{4}\left(\sigma^2+\bbox{\pi}^2\right)^2 \nonumber \\ [0.3cm]
&-&\bar{\psi}\left[g_{0\omega}\gamma_{\mu}\omega^{\mu}\right]\psi
-\frac{1}{4}F_{\mu\nu}F^{\mu\nu}+\frac{1}{2}m_\omega^2\omega_{\mu}
\omega^{\mu} \label{lag} \\ [0.6cm]
{\cal L}_b&=&\left(\epsilon c-vm_\pi^2\right)s\;.
\end{eqnarray}
\noindent 
where we have used the abbreviations:
\begin{eqnarray}
M&=&gv \;,\label{NT1}\\ [0.3cm]
m_\pi^2&=&\mu_0^2+\lambda^2v^2 \;, \\ [0.3cm]
m_\sigma^2&=&\mu_0^2+3\lambda^2v^2 \;.
\end{eqnarray}
\noindent
In the form of Eq.~(\ref{lag}), the nucleon mass is M and is given by
(\ref{NT1}), which is the Goldberger-Treiman relation for $g_A=1$; the pions
and the $\sigma$ are no longer degenerate in mass.
The value of $v$ is determined by the condition (\ref{Eq.s}) and the 
PCAC relation, which gives $v=f_\pi$ at tree level, $f_\pi$ being the 
pion decay constant. As pointed out by Lee
\cite{Lee72}, both lagrangians (\ref{lags}) and (\ref{lag}) are completely
equivalent. All the Feynman rules derived from this lagrangian are very 
straightforward and more details can be found on Lee's book.

Next, we discuss the spectral representations of the nucleon propagator and of
its inverse. We do not intend to review the subject of spectral representations,
we simply make use of the relevant equations for the purposes of the present
paper. We refer the reader to Refs.~\cite{Sch62,Rom69,BLT75}. The nucleon
propagator is defined as
\begin{equation}
G_{\alpha \beta}(x'-x)=-i<0|T[\psi_{\alpha}(x')\bar \psi_{\beta}(x)]|0>\;,
\label{defnucpro}
\end{equation}
where $|0\!>$ represents the physical vacuum state. Following the BPW 
approach, the K\"allen-Lehmann representation for the Fourier
transform $G(p)$ of $G(x'-x)$ can be written as
\begin{equation}
G(p)=\int_{- \infty}^{+ \infty} d\kappa\; {A(\kappa) \over {{\not\!p} - \kappa
+ i\epsilon}}\;,
\label{kaleh} 
\end{equation}
where $A(\kappa)$ is the spectral function.  It represents the probability that
a state of mass $|\kappa|$ is created by $\psi$ or $\bar \psi$, and as such it
must be non-negative. Negative $\kappa$ corresponds to states with opposite
parity to the nucleon.

Defining the projection operators
\begin{equation}
P_{\pm}(p)={1\over 2}\left(1 \pm {{\not\!p} \over w_p}\right)\;,
\end{equation}
where
\begin{eqnarray}
w_p = \sqrt{p^2} =
            \left\{ \begin{array}{ll}
                           \sqrt{p^2}, & \mbox{if $p^2 > 0$} \\
                          i\sqrt{-p^2}, & \mbox{if $p^2 < 0$},
                                        \end{array}
                                        \right.
\label{wp}
\end{eqnarray}
$G(p)$ can be rewritten conveniently as
\begin{eqnarray}
G(p)=P_{+}(p)\tilde G (w_p+i\epsilon)+P_{-}(p)\tilde G(-w_p-i\epsilon)\;,
\label{gpro}
\end{eqnarray}
where $\tilde G(z)$
is given by the dispersion integral
\begin{equation}
\tilde G(z) = \int_{- \infty}^{+ \infty} d\kappa\;
{A(\kappa) \over {z - \kappa}}\;.
\label{gtil}
\end{equation}

The inverse of the propagator can also be written in terms of the projection 
operators $P_{\pm}(p)$ as
\begin{equation}
G^{-1}(p)=P_{+}(p)\tilde G^{-1} (w_p+i\epsilon)+
P_{-}(p)\tilde G^{-1}(-w_p-i\epsilon)\;.
\label{ginvpro}
\end{equation}
The spectral representation for $\tilde G^{-1}(z)$ is written as,
\begin{eqnarray}
\tilde G^{-1}(z)&=&z-M_0 - \tilde \Sigma(z)\nonumber \\ [0.3cm]
&=&z-M_0 - \int_{- \infty}^{+ \infty} d\kappa
{T(\kappa) \over {z-\kappa}}\;.
\label{nucself}
\end{eqnarray}
The function $\tilde \Sigma(z)$ is related to the nucleon self-energy 
$\Sigma(q)$  (see Eq. (\ref{sdenuc})) by the projection operators 
$P_{\pm}(q)$ as in Eq. (\ref{ginvpro}).

Since $A(\kappa)$ is supposed to be non-negative, $\tilde G(z)$
and $\tilde G^{-1}(z)$ can have no
poles or zeros off the real axis. This is known as the Herglotz property 
\cite{BPW}.  In general, the integral in Eq. (\ref{nucself}) needs
renormalization. The usual mass and wave-function renormalizations are 
performed by imposing the condition that the renormalized propagator has a pole
at the physical nucleon mass $M$, with unit residue. This implies that 
the renormalized propagator $\tilde G_{R}(z)$, defined as

\begin{equation}
\tilde G_{R}(z) \equiv \tilde G(z)/ Z_2\;,
\label{defgr}
\end{equation}
is given by the following expression:
\begin{equation}
\tilde G_R(z) = \int_{- \infty}^{+\infty} d\kappa\; {A_{R}(\kappa) \over
{z-\kappa}}\;,
\label{rg}
\end{equation}
and the inverse of the propagator is given by
\begin{eqnarray}
&&\tilde G_{R}^{-1}(z) =(z-M) \times \nonumber \\ [0.3cm]
&&\left[ 1-(z-M)\int_{- \infty}^{+\infty} d\kappa\;
{T_{R}(\kappa) \over {(\kappa-M)^2(z-\kappa)}}\right]\;.
\label{rself}
\end{eqnarray}
In the above expressions, $A_{R}(\kappa) = A(\kappa)/Z_2$ and 
$T_{R}(\kappa) = Z_2T(\kappa)$. In terms of renormalized quantities, $Z_2$ can
be written as
\begin{equation}
Z_2 = 1 - \int_{- \infty}^{+\infty} d\kappa {T_{R}(\kappa)
\over{(\kappa-M)^2}}
\label{z2t}
\end{equation}
or
\begin{equation}
Z_2=\left[\int_{- \infty}^{+\infty} d\kappa\; A_{R}(\kappa)\right]^{-1}\;.
\label{z2a}
\end{equation}

The spectral functions $A_R(\kappa)$ and $T_R(\kappa)$ are related by 

\begin{eqnarray}
A_R(\kappa)&=&\delta(\kappa-M)+
|{\tilde G}_R^{-1}(\kappa(1+i\epsilon))|^{-2}T_R(\kappa)\label{AandT} \\ [0.3cm]
&\equiv &\delta(\kappa-M)+\bar A_R(\kappa)\;.
\label{Abar}
\end{eqnarray}

\subsection{Schwinger-Dyson equation for the nucleon propagator}
 
\begin{figure}
\vspace{0.5cm}
\epsfxsize=8cm
\epsfysize=4cm
\centerline{\epsffile{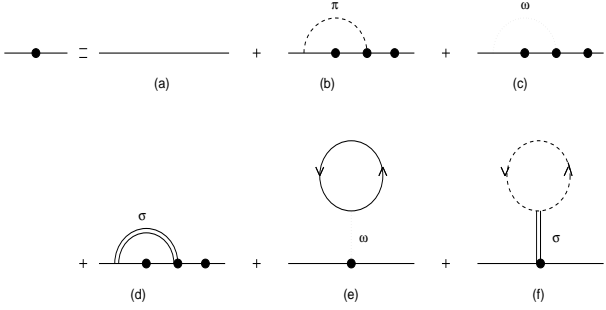}}
\vspace{0.5cm}
\caption{Diagrammatic representation of SDE. The tadpole diagrams do not
contribute because they drop out in the renormalization process.}
\label{Fig.2}
\end{figure}

In this section we calculate the ``dressed'' nucleon propagator with its 
self-energy given by contributions of $\pi$, $\omega$, and $\sigma$ mesons. 
The Schwinger-Dyson equation (SDE) for the nucleon propagator in momentum 
space is given by the following expressions
\begin{equation}
G(p)=G^{(0)}(p)+G^{(0)}(p)\Sigma(p)G(p),
\label{sdenuc}
\end{equation}
where
\begin{eqnarray}
\Sigma(p)&=&- i g^2_0 \int {d^4q\over (2\pi)^4} \gamma_5 \tau^i
D_{\pi}(q^2)G(p-q)\Gamma_5^i(p-q,p;q) \nonumber \\ [0.3cm]
&+&i g_{0\omega}^2 \int {d^4q\over (2\pi)^4} \gamma_{\mu}
D_{\omega}^{\mu \nu}(q^2)G(p-q)\Gamma_{\nu}(p-q,p;q)\nonumber \\ [0.3cm]
&+&i g^2_0 \int {d^4q\over (2\pi)^4} D_{\sigma}(q^2)G(p-q)\Gamma_{S}(p-q,p;q),
\label{nuceq}
\end{eqnarray}
is the nucleon self-energy, shown schematically in Fig.~\ref{Fig.2}. 
%
%
In
Eq.~(\ref{nuceq}), $D_\pi$, $D_\omega^{\mu\nu}$, and $D_\sigma$ are the $\pi$, 
$\omega$, and $\sigma$ propagators and $\Gamma_5^i(p-q,p;q)$, 
$\Gamma_\nu (p-q,p;q)$, and $\Gamma_S(p-q,p;q)$  are respectively the
pion-nucleon, omega-nucleon, and sigma-nucleon vertex functions.
The tadpoles in Fig.~\ref{Fig.2} do not contribute to the nucleon propagator,
since they drop out in the renormalization procedure. 
The Hartree-Fock (HF) approximation amounts to using the noninteracting 
meson propagators and bare vertices $\Gamma_5^j(p-q,p;q)=\tau^i\gamma_5$,
$\Gamma_\nu (p-q,p;q)=\gamma_\nu$, and $\Gamma_S(p-q,p;q)=1$ 
in Eq.~(\ref{nuceq}). In order to regulate the ultraviolet behavior of 
the interaction and study the role of a ghost-free self-consistent propagator,
we consider simplified vertex functions that are written as:
\begin{eqnarray}
\Gamma_5^i(p_1, p_2; q)&=&\tau^i \gamma_5 F_5(p_1, p_2; q) \label{Gam5} \\ [0.3cm]
\Gamma^{\mu}(p_1, p_2; q)&=&\gamma^{\mu} F_V(p_1, p_2; q) \label{GamV}\\ [0.3cm]
\Gamma_S(p_1,p_2;q)&=&F_S(p_1,p_2;q) \label{GamS}\;,
\end{eqnarray}                                               
where $F_5(p_1, p_2; q)$, $F_V(p_1, p_2; q)$, and $F_S(p_1,p_2;q)$
are scalar functions. Substituting Eqs.~(\ref{Gam5}-\ref{GamS}) 
and the spectral representations for $G(p)$ in the integral for $\Sigma(p)$, 
Eq.~(\ref{nuceq}), and using the projection operators $P_{\pm}(p)$, one 
obtains:
\begin{equation}
T_R(\kappa)=\int_{-\infty}^{+\infty} d\kappa'K(\kappa,\kappa')A_R(\kappa')\;,
\label{tkap}
\end{equation}
where $K(\kappa,\kappa')$ is the scattering kernel given by
\begin{eqnarray}
K(\kappa, \kappa') &=& K_{\pi}(\kappa, \kappa'; m_{\pi}^2) \nonumber \\ [0.3cm]
&+&2 K_{\omega}(\kappa, \kappa'; m_{\omega}^2) +
K_{\sigma}(\kappa, \kappa'; m_{\sigma}^2)\;;
\label{K}
\end{eqnarray}
with $K_{\pi}(\kappa, \kappa'; m^2_\pi)$, $K_{\omega}(\kappa, \kappa'; 
m^2_\omega)$, and $K_{\sigma}(\kappa, \kappa'; m^2_\sigma)$ being
respectively the $\pi$, $\omega$, and $\sigma$ contributions, given by
\begin{eqnarray}
&&K_{\pi}(\kappa, \kappa'; m^2_\pi )= F_5(\kappa, \kappa'; m_\pi)
\; 3\,\left({g\over 4\pi}\right)^2 \;{1\over 2|\kappa|^3} 
\nonumber \\ [0.3cm]
&\times&\left[\kappa^4-2\kappa^2({\kappa'}^2+m^2_\pi )+({\kappa'}^2
-m^2_\pi )^2\right]^{1/2}\nonumber \\ [0.3cm]
&\times&\left[(\kappa-\kappa')^2-m^2_\pi \right]\;
\theta(\kappa^2-(|\kappa'|+m_\pi)^2)\;,
\label{Kpi}
\end{eqnarray}
\begin{eqnarray}
&&K_{\omega}(\kappa, \kappa'; m_\omega^2)= F_V(\kappa, \kappa'; m_\omega)\;
\left({g_{\omega}\over 4\pi}\right)^2\;{1\over 2|\kappa|^3}
\nonumber \\ [0.3cm]
&\times&\left[\kappa^4-2\kappa^2({\kappa'}^2+m_\omega^2)+({\kappa'}^2
-m_\omega^2)^2\right]^{1/2}\nonumber \\ [0.3cm]
&\times&\left[(\kappa-\kappa')^2-2\kappa \kappa'-m_\omega^2\right]
\;\theta(\kappa^2-(|\kappa'|+m_\omega)^2)\;,
\label{Komega}
\end{eqnarray}
\noindent and 
\begin{eqnarray}
&&K_{\sigma}(\kappa, \kappa'; m^2_\sigma )=F_S(\kappa, \kappa'; m_\sigma)\;
\left({g\over 4\pi}\right)^2\;{1\over 2|\kappa|^3}\;
\nonumber \\ [0.3cm]
&\times&\left[\kappa^4-2\kappa^2({\kappa'}^2+m^2_\sigma )+({\kappa'}^2
-m^2_\sigma )^2\right]^{1/2}\nonumber \\
&\times&\left[(\kappa+\kappa')^2-m^2_\sigma \right]
\theta(\kappa^2-(|\kappa'|+m_\sigma)^2)\;.
\label{Ksigma}
\end{eqnarray}
In the above equations, $g$ and $g_\omega$ are the renormalized coupling
constants, defined as $g=Z_2g_0$ and $g_\omega=Z_2 g_{0\omega}$.


\subsection{Ghost poles}

Next we discuss the numerical solution of the SDE. The problem consists
in solving for the spectral function $A_R(\kappa)$. The equations involved 
are  Eqs. (\ref{rself}), (\ref{Abar}) and (\ref{tkap})-(\ref{Ksigma}). These
represent a set of coupled nonlinear integral equations which are solved by
iteration~\cite{BPW,Kre+93}.

Initially, we consider bare vertices: $F_5(p_1, p_2, q)=F_V(p_1, p_2, q)=
F_S(p_1, p_2,q)=1$, and study the convergence properties of the SDE for the
nucleon self-energy. The new aspect here is the presence of the chiral partner
of the pion, the $\sigma$. We consider the following cases: 
(a) $\sigma$ meson only,
(b) $\pi+\sigma$ mesons, and (c) $\pi+\omega+\sigma$ mesons. We use the
following values for the coupling constants:
\begin{eqnarray}
{g^2_{\pi}\over 4\pi}=\frac{g^2_\sigma}{4\pi} \equiv \frac{g^2}{4\pi}&=&
14.6 \\ [0.3cm]
{g_{\omega}^2 \over 4\pi}&=&6.36 \;,
\label{couplings}
\end{eqnarray}
where we wrote explicitly that the value of the sigma-nucleon coupling constant
is equal to the pion-nucleon one, as required by the linear realization of
chiral symmetry.

\begin{figure}
\epsfxsize=7.2cm
\epsfysize=9cm
\centerline{\epsffile{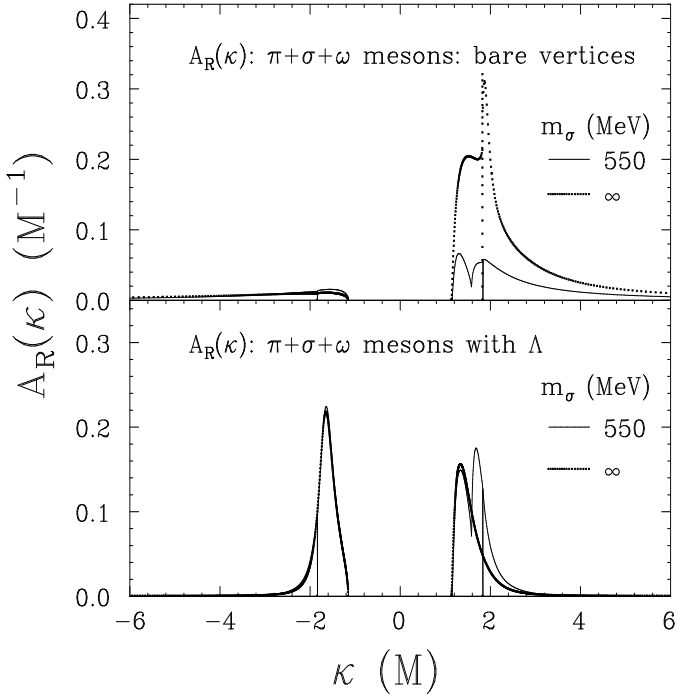}}
\vspace{1.0cm}
\caption{Top: plot of $A_R(\kappa)$ as a function of $\kappa$
for nucleon self-energy due to $\pi+\sigma+\omega$ mesons with {\em bare}
nucleon vertices. Bottom: the same study for the same system but with 
vertices dressed by form factors. For the other $\sigma$ masses, the  
curves are between these two curves.}
\label{Fig.4}
\end{figure}

The first fact we observed in solving the SDE was that the introduction of the
chiral partner of the pion {\em does not remove} the ghost poles.
As the mass of the $\sigma$ meson remains a point of debate, we varied 
$m_\sigma$ over a wide range. The solutions of SDE converge quickly in the
studied range, $500\leq m_\sigma \leq 1500$ MeV. For the case of using the
$\sigma$ only , the convergence is more difficult to achieve for $m_\sigma$
between 550 and 770 MeV, but it was obtained by using a small convergence 
factor at each iteration. However, 
this case is of little interest for $\pi N$ scattering, since pions are
always present. The converged spectral functions $A_R(\kappa)$ are shown
in Fig.~\ref{Fig.4}. We used the following values for the meson masses:
\begin{eqnarray}
m_\pi&=&138.03\mbox{ MeV}\;, \\ [0.2cm]
m_\omega&=&783\mbox{ MeV}\;,\\ [0.2cm]
m_\sigma&=&550,\;770,\;980\mbox{ MeV and }m_\sigma\rightarrow
\infty\;.
\label{msigma}
\end{eqnarray}
The reason for using this particular set of $\sigma$ masses is the following: 
$m_\sigma=550$ MeV is the value commonly used in the One Boson 
Exchange Potentials (OBEP)~\cite{NN,MHE}; $m_\sigma=770$ MeV is the 
value used by  Serot and Walecka~\cite{SW86} in the calculations of nuclear
matter properties using the chiral linear sigma model; $m_\sigma=980$ MeV
is the first scalar meson in the mesons table, $f_0$, and the limit
$m_\sigma\rightarrow\infty$ supplies the connection between the linear
realization of chiral symmetry and the minimal chiral model of the non-linear
realization of chiral symmetry in $\pi N$ system \cite{RR94}. Recently, 
T\"ornqvist and Roos \cite{TR94} claimed that the sigma meson really exists,
with a mass of 860 MeV and an extremely broad width of 880 MeV. Although
this fact is receiving great attention lately, a final word of 
$\sigma$ existence in this mass region remains to be stated. 

Fig.~\ref{Fig.4}-top presents the nucleon dressed by the $\pi+\sigma+\omega$
mesons, for $m_\sigma=550$ MeV and $m_\sigma\rightarrow\infty$. We note that 
$A_R(\kappa)$ for positive $\kappa$ is much larger than for negative $\kappa$,
it increases as the $\sigma$ mass increases, and becomes equal to the 
$\pi +\omega$ case (NW's study) in the limit $m_\sigma\rightarrow\infty$. Recall
that for $m_\sigma\rightarrow\infty$ the $\sigma$ meson does not contribute to
$A_R(\kappa)$. Starting from $m_\sigma\rightarrow\infty$ and going down, one
finds that the presence of the $\sigma$ meson modifies drastically the
spectral function for positive $\kappa$, decreasing the main peak and creating
a second resonance peak; for negative $\kappa$, the changes are very small. 
The zeros of $A_R(\kappa)$ at $\kappa=\pm(M+m_\omega)$ shown in the 
figure are due to the discontinuity of the $\omega$ kernel, Eq.~(\ref{Komega}),
at these points as explained in Ref.~\cite{Kre+93}. The $\pi+\sigma$ system 
revealed minor differences, since the contribution of
the $\omega$ meson is small and come from the region 
near $\kappa=\pm(M+m_\omega)$. The position of the ghosts poles and the value
of their residues are shown in Tab.~\ref{Tab.01}, where one sees that the 
real part of the complex pole position ($P_R$) increases with the 
$\sigma$ mass. At the usual $m_\sigma=550$ MeV, $P_R$ is a little below
$2m_\pi$, which means that the inclusion of the $\sigma$ meson in the nucleon
self energy brings down the position of the ghosts from 1 GeV to near
300 MeV. 

\begin{table}
\caption{Ghost poles positions and residues for $\pi+\sigma+\omega$ dressing
with different $\sigma$ masses.}
\begin{tabular}{lcr}
Process & Pole position                   & Pole residue \\ 
        & (units of M)                    & (dimensionless) \\ \tableline
$\pi+\omega+\sigma(550)$
          & $0.2624\pm i\;0.4402$         & $-0.6208\pm i\;1.1281$ \\
$\pi+\omega+\sigma(770)$
          & $0.4043\pm i\;0.7648$         & $-0.6317\pm i\;0.5616$ \\
$\pi+\omega+\sigma(980)$
          & $0.5147\pm i\;0.9147$         & $-0.6431\pm i\;0.4143$ \\
$\pi+\omega+\sigma(\infty)$
          & $1.1058\pm i\;1.1337$         & $-0.7410\pm i\;0.1788$ 
\end{tabular} 
\label{Tab.01}
\end{table}

Next we consider vertex form factors. In principle one can use the
corresponding Sudakov form factors, as in Ref.~\cite{Kre+93}. However, since
the Sudakov form factor is known exactly in the ultraviolet only, one has
to interpolate it in some way down to the infrared or simply parametrize its
infrared behavior. However, since for our purposes here of killing the
ghosts the infrared behavior is not relevant, we prefer to simplify matters
and use parametrized form factors, which implies that we can not make any 
conclusion involving chiral symmetry in these approach, 
since this phenomenological vertex
dressing has no connection with chiral symmetry. 

For general off-shell legs, we use the
factorized form of Pearce and Jennings~\cite{PJ91}; for a vertex with 
four-momenta $p_\alpha$, $p_\beta$, $p_\gamma$, the form factor is
\begin{equation}
F_{\alpha\beta\gamma}=F_\alpha(p_\alpha^2)F_\beta(p_\beta^2)F_\gamma
(p_\gamma^2)\;.
\end{equation}
For mesons we adopt the expressions by Gross, Van Orden and Holinde~\cite{GVH92}
\begin{equation}
F_m(q^2)=\left[\frac{1+\left(1-\mu_m^2/\Lambda_m^2\right)^2}
{1+\left(1-q^2/\Lambda_m^2\right)^2}\right]^2
\label{ffm}
\end{equation}
where $\Lambda_m$ is the meson cutoff mass. For the nucleon legs we adopt
the expressions by Gross and Surya~\cite{GS93}
\begin{equation}
F_B(p^2)=\frac{(\Lambda_B^2-m_B^2)^2}{(\Lambda_B^2-m_B^2)^2+(m_B^2-p^2)^2}
\label{ffp}
\end{equation}
\noindent where $\Lambda_B$ is the nucleon cutoff mass. Both meson 
and nucleon form factors have the correct on-shell limit, equal to unity.

At this point it is perhaps convenient to call attention that we use form
factors for regulating the ultraviolet with the only aim of studying the role 
of a ghost-free propagator in $\pi N$ scattering. In principle, the form 
factors are calculable within the model by means of vertex corrections. In 
particular, such vertex corrections must satisfy Ward-Takahashi identities 
that follow from chiral symmetry, and of course our form factors $F_5$ 
and $F_S$, Eq.~(\ref{ffm}), do not satisfy such identities. This interesting 
subject is intended to be pursued in a future work.

The cutoff values $\Lambda_B$ and $\Lambda_m$ are constrained to kill the ghost
poles and  give the best fit to the scattering lengths. There is a critical
value $\Lambda_c$ such that for $\Lambda < \Lambda_c$ the ghosts 
disappear~\cite{Kre+93}. The values for $\Lambda_c$ for all systems studied in
this paper are shown in the second column of Tab.~\ref{Tab.03}. Note that the
form factors corresponding to the meson legs do not contribute in the SDE
(see Eqs.~(\ref{Kpi}-\ref{Ksigma})). One observes that $\Lambda_B < \Lambda_c$. 
The constraints related to the scattering lengths
are discussed in Sec.~\ref{3.3}.

Fig.~\ref{Fig.4}-bottom presents the case $\pi+\sigma+\omega$, with form 
factors at each vertex. We use $\Lambda_B=1330$ MeV. The shape of $A_R(\kappa)$
depends strongly on the $\sigma$ mass for $\kappa>0$ only. One sees that the
second peak is mainly due to the $\sigma$ meson, it decreases as the $\sigma$
mass increases and disappears in the limit $m_\sigma\rightarrow\infty$. 
The interesting effect due to the form factors is that the spectral function 
for $\kappa < 0$ becomes very large as compared to the case of bare vertices.
This will have serious consequences for the observables of $\pi N$ scattering.
The position of the peaks in the spectral function for the different cases
studied are presented in the third to fifth columns of Tab.~\ref{Tab.03}. 

\begin{table}
\caption{Critical values of the nucleon cutoff (in {\rm MeV}), position of
the peaks in $A_R(\kappa)$ (in {\rm MeV}), and the integral of $A_R(\kappa)$
over $\kappa$.}
\begin{tabular}{lccccr}
System & $\Lambda_c$ & $1^{st}$ Peak & $2^{nd}$ Peak & 
$3^{rd}$ Peak & Area\\ \tableline
$\pi$ only                  & 2085 & -1541 & 1263 & -    & 0.1923 \\
$\omega$ only               & 9550 & -1807 & 1777 & -    &-0.0008 \\
$\pi+\omega$                & 1815 & -1536 & 1268 & -    & 0.1919 \\
$\sigma(550)$               & 2650 & -1717 & 1569 & -    & 0.0259 \\
$\sigma(770)$               & 3477 & -1967 & 1778 & -    & 0.0045 \\
$\sigma(980)$               & 4523 & -2208 & 1997 & -    & 0.0011\\
$\sigma(1581)$              & 5390 & -2913 & 2628 & -    & 0.0001\\ \tableline
$\pi+\sigma(550)$           & 1817 & -1545 & 1257 & 1581 & 0.2328 \\
$\pi+\sigma(770)$           & 1951 & -1541 & 1262 & 1751 & 0.2002 \\
$\pi+\sigma(980)$           & 2026 & -1541 & 1263 & 1933 & 0.1947 \\
$\pi+\sigma(\infty)$        & 2071 & -1541 & 1265 &  -   & 0.1923 \\ \tableline
$\pi+\omega+\sigma(550)$    & 1822 & -1541 & 1257 & 1581 & 0.2324 \\
$\pi+\omega+\sigma(770)$    & 1958 & -1536 & 1263 & 1723 & 0.1998\\
$\pi+\omega+\sigma(980)$    & 2037 & -1536 & 1268 & 1940 & 0.1943\\
$\pi+\omega+\sigma(\infty)$ & 2085 & -1541 & 1263 & -    & 0.1923
\end{tabular}
\label{Tab.03}
\end{table}

The contribution of each meson to the nucleon self-energy can be estimated by 
the integral over the spectral function $A_R(\kappa)$. The integral is related 
to the renormalization constant $Z_2$ as indicated in Eq.~(\ref{z2a}). This
is shown in the last column of Tab.~\ref{Tab.03}. One notices that the
pion gives the highest contribution, followed by the lightest sigma
meson (550 MeV) and the omega meson. 
This indicates that the dressing of the nucleon is mainly due to the pion.

\section{$\pi N$ scattering and ghost poles}
\label{sec:interplay}

\subsection{Introduction}

\begin{figure}
\vspace{0.5cm}
\epsfxsize=8cm
\epsfysize=3cm
\centerline{\epsffile{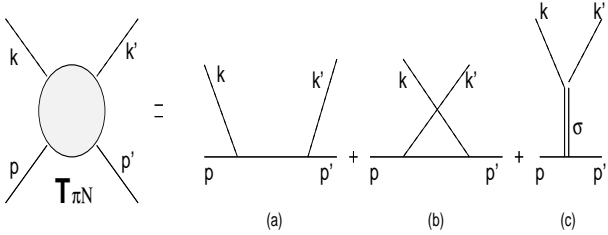}}
\vspace{0.6cm}
\caption{(a),(b) Order $g^2$ contributions involving only pions and
nucleons. (c) Scalar meson contribution.}
\label{Fig.6}
\end{figure}

The simplest field-theoretical model for $\pi N$ scattering is the summation 
of Feynman diagrams in tree approximation~\cite{GLM93}. Such a model can
involve only pions and nucleons~\cite{GSS88} or it may be augmented by
hadronic resonances like the $\Delta$ and the Roper in the baryonic sector or 
the $\sigma$, $\rho$ and others mesons in the mesonic sector~\cite{GLM93}. 
The differences between these models come from the inclusion or not of chiral
symmetry and how far one desires to reproduce the experimental data. In Born
approximation, when using pions and nucleons only, the lowest order 
contribution is the sum of just two graphs, as shown in Figs.~\ref{Fig.6}(a,b).
As is well known, these first two contributions give bad results for
isoscalar observables.

According to the $\sigma$-linear lagrangian, Eq.~(\ref{lag}),
the $\pi N$ coupling  is pseudoscalar (PS) and 
two new couplings appear, $\sigma NN$ and $\pi\pi\sigma$. The lowest order 
tree diagrams contain one more diagram, as shown in
Fig.~\ref{Fig.6}(c), which inclusion led to an almost perfect fit to 
the value of the isospin-even amplitude $A^{(+)}$ at threshold.
This is one of the classical examples of the importance of chiral symmetry in
hadronic interactions. 

In the 70's, Nutt and Wilets~\cite{nutwil} found the result that the  
threshold value of $A^{(+)}$ could be explained by introducing 
quantum fluctuations in the nucleon propagator. 
NW used the BPW formalism to solve the
SDE equation for the nucleon propagator. As discussed in 
the previous section, the renormalized nucleon propagator is well defined 
and self-consistent, but contains a pair of complex ghost poles. 
In the Sec.~\ref{3.3} we re-investigate this result on the light
of dressed nucleon propagators free of ghosts.

\subsection{Bare nucleons}

At lowest order (tree level) the scattering amplitude for 
$\pi^\alpha(k)N(p)\longrightarrow \pi^\beta(k')N(p')$, where $p$ and $p'$ 
($k$ and $k'$ ) are the nucleon (pion) four-momentum, and $\alpha$ and $\beta$
the isospin labels, can be described by a model involving the
processes shown in Fig.~\ref{Fig.6}, plus the contribution of $\Delta$  and
$\rho$ resonances. Such a model provides a good description of data from
threshold until pion energies up to 350 MeV~\cite{OO75}, 
400 MeV~\cite{PJ91}, and 600 MeV~\cite{GS93}. The $\sigma$-term is needed for
 the implementation of chiral symmetry in a linear way. At  tree level, 
the $\sigma$-exchange can be understood as representing a function that, in
the context of current algebra, comes from the equal-time commutator of an 
axial current and its divergence. Usually this contribution is represented by 
a parameterized form~\cite{Coo+79,CG81}. 

The $\pi N$ amplitude $T_{\pi N}$ can be parameterized as~\cite{nutwil}
\begin{eqnarray}
T_{\pi N} &=& \bar{u}\,({\bf p}')\left\{\left[{A}^+ +
   \frac{1}{2}\,(\not\! k + \not\! k^\prime) B^+\right] 
   \delta_{ab} \right. \nonumber \\ [0.3cm]
&+&\left.\left[A^- +\frac{1}{2}\,
  (\not\! k + \not\! k') 
  B^-\right] i \epsilon_{bac}\, \tau_c\right\}u({\bf p})\;,
\label{tamp}
\end{eqnarray}
where $A^\pm$ and $B^\pm$ are Lorentz invariant functions that contain the
dynamics of the model. Let us first examine the nucleon pole and scalar
exchange. For the diagrams shown in Fig.~\ref{Fig.6} we get
\begin{eqnarray}
A^+(s,t,u)&=&\frac{g^2}{M} \,\left[\frac{1-(m_\pi/m_\sigma)^2}
{1-t/m_\sigma^2}\right], \label{A+}\\ [0.3cm] 
A^-(s,t,u)&=&0,\label{A-} \\ [0.3cm]
B^+(s,t,u)&=&-g^2\, \left[\frac{1}{s-M^2}-\frac{1}{u-M^2}\right],\label{B+} \\
[0.3cm] 
B^-(s,t,u)&=&-g^2\, \left[\frac{1}{s-M^2}+\frac{1}{u-M^2}\right]\label{B-}.
\end{eqnarray}
$B^+$ and $B^-$ receive contributions from nucleon intermediate states, and 
$A^+$ from $\sigma$ exchange. If one takes the limit 
$m_\sigma\rightarrow\infty$ for a fixed $t$, the $A^+$ result is 
\begin{equation}
A^+(s,t,u)=\frac{g^2}{M}.
\end{equation}
This limit corresponds to a contact scalar interaction of two pions with the
nucleon, and is the exact result of Adler theorem~\cite{Adl65,Wei66}.  
It is worth noting that the isospin even exchange term ($A^+ + m_\pi\,B^+$) 
vanishes in the zero four momentum transfer limit at threshold 
($s=M+m_\pi,\,t=0,\,u=M-m_\pi$), if we neglect terms proportional to
${m_\pi^2}/{M^2}$.

Next we calculate the scattering length $a$. It is given in terms of the
T-matrix as: 
\begin{equation}
a=\frac{1}{8\pi(M+m_\pi)}|T|_{\text{\scriptsize threshold}},
\label{a}
\end{equation}
where $|T|_{\text{\scriptsize threshold}}$ is the scattering amplitude 
$T$ calculated at
threshold ($p=M$). $T$ is related to the differential cross section as
\begin{equation}
\frac{d\sigma}{d\Omega}=\frac{1}{(8\pi W)^2}\, |T|^2,
\end{equation}
where $W$ is the total invariant mass of the final state. In deriving
Eq.~(\ref{a}) the cross section was approximated  by the area of a black sphere
of radius $a$, which is a good approximation at very low energies. In terms of
the invariants $A^\pm$ and $B^\pm$, $a$ can be written as (in the c.m. frame):
\begin{equation}
a^\pm=\frac{1}{4\pi\left(1+m_\pi/M\right)}\cdot\left(A^\pm+m_\pi B^\pm
\right).
\label{Eq.1}
\end{equation}
        
The experimental values for the scattering lengths are~\cite{OO75}
\begin{eqnarray}
a^+&=&-(0.021\pm0.021) \mbox{ fm} \nonumber \\ [0.3cm]
a^-&=&0.139 \left\{\begin{array}{ll}
                           +0.004 & \mbox{ fm} \\
                           -0.010 & \mbox{ fm}
                                        \end{array}
\right.
\label{Eq.2}
\end{eqnarray}
In Tab.~\ref{Tab.04} we show the results using the delta function piece of
the nucleon propagator with and without chiral symmetry (CS). In obtaining 
these results no form-factors at the meson-nucleon vertices were used. One 
sees that inclusion of chiral symmetry improves the results for the scattering
lengths in the isoscalar channel. This is due to a strong
cancellation of the nucleon pole \hfill

\onecolumn
($B^+$) with the $\sigma$ exchange ($A^+$).
This cancellation is exact if we take the limit $m_\sigma\rightarrow\infty$ 
and $\left({m_\pi^2}/{M^2}\right) \rightarrow 0$. The magnitude of the
scattering length $a^+$ is almost zero, it is proportional to ${m_\pi^2}/{M^2}$.
In the chiral limit, $m_\pi\rightarrow 0$, the scattering lengths are zero.
This is a property of the linear $\sigma$ model lagrangian at the tree 
level.

\begin{table}
\caption{Scattering lengths with and without chiral symmetry at the tree 
level, using bare nucleon propagator and bare vertices. Chiral symmetry is 
indicated by C.S. The experimental values are given in
Eq.~(\protect\ref{Eq.2}).}
\begin{tabular}{lcccccr}
Process & $A^+$   & $A^-$ & $m_\pi. B^+$ & $m_\pi. B^-$ & $a^+$  & $a^-$ \\ 
&($g^2/M$)& ($g^2/M$)  & ($g^2/M$)    & ($g^2/M$)         & (fm)   & (fm)  \\ 
\tableline
Fig.~\ref{Fig.6}(a+b) - no C.S.
          & 0       & 0     & -1.005433    & 0.07391    & -2.6896 & 0.1977  \\ 
\tableline
Fig.~\ref{Fig.6}(a+b+c) - with C.S. \\
$m_\sigma=550$ MeV
          & 0.93700 & 0     & -1.005433    & 0.07391    & -0.1830  & 0.1977 \\
$m_\sigma=770$ MeV
          & 0.96786 & 0     & -1.005433    & 0.07931    & -0.1005  & 0.1977 \\
$m_\sigma=980$ MeV
          & 0.98020 & 0     & -1.005433    & 0.07931    & -0.0676  & 0.1977 \\
$m_\sigma\rightarrow\infty$
          & 1.0000 & 0      & -1.005433    & 0.07931    & -0.0145  & 0.1977
\end{tabular}
\label{Tab.04}
\end{table}

\subsection{Dressed nucleons}
\label{3.3}

The graphs that contribute to the $\pi N$ scattering are given in 
Fig.~\ref{Fig.6}, where the nucleon propagators are now 
``dressed'' by the $\pi+\sigma+\omega$ mesons. The new contributions, as we 
compared to the NW's work, are the inclusion of the scalar meson both in 
$t$ channel scattering and in the nucleon self energy. The contribution of the
non-delta part of the spectral function, defined in
Eq.~(\ref{Abar}), to the functions $A^\pm$, $B^\pm$ is given by:
\begin{eqnarray} 
A^\pm(s,t,u)&=&g^2\int_{-\infty}^\infty d\kappa\,\bar A(\kappa)\,
(\kappa-M)\, \left[\frac{1}{s-\kappa^2}\pm\frac{1}{u-\kappa^2}\right]  
\label{A1} \\ [0.3cm]
B^\pm(s,t,u)&=&g^2\int_{-\infty}^\infty d\kappa\, \bar A(\kappa)\, \left[
-\frac{1}{s-\kappa^2}\pm\frac{1}{u-\kappa^2}\right] 
\label{A3}
\end{eqnarray}
The total contribution for $A^\pm$ and $B^\pm$ is the sum of three parts
coming from: (a) the delta function part of $A(\kappa)$, given in 
Eqs.~(\ref{A+}-\ref{B-}), (b) the self-energy given by Eqs.~(\ref{A1}-\ref{A3}),
and (c) the ghost poles. The contribution from the ghosts poles is given by:
\begin{eqnarray} 
A^\pm_g(s,t,u)&=&g^2\sum_c A_c\, (\kappa_c-M)\,
\left[\frac{1}{s-\kappa_c^2}\pm\frac{1}{u-\kappa_c^2}\right]\label{AC1} 
\\ [0.3cm]
B^\pm_g(s,t,u)&=&g^2\sum_c A_c\,\left[-\frac{1}{s-\kappa_c^2} \pm
\frac{1}{u-\kappa_c^2}\right], 
\label{AC2}
\end{eqnarray}
where the sum is over ($\kappa_c,A_c$) and its complex conjugate
($\kappa_c^*,A_c^*$). 

\begin{table}
\caption{Results for observables using dressed nucleon propagators: (a)
$m_\pi=$138.08 MeV and (b) the chiral limit $m_\pi\rightarrow 0$. No form factors are used.}
\begin{tabular}{l|ccc|rrr} 
             &\multicolumn{3}{c|}{(a) $m_\pi=$138.08 MeV}            
             &\multicolumn{3}{c}{(b) $m_\pi=0$}  \\ \tableline
Contribution & $A^+$  & $a^+$              & $a^-$ & $A^+$ & $a^+$     & $a^-$
\\ 
& ($g^2/M$)& (fm) & (fm)    & ($g^2/M$)       & (fm)   & (fm)   \\ 
\tableline
1) $\pi+\sigma(550)+\omega$: $\bar{A}_R(\kappa)$ 
        &  0.0197  &  0.0773  & 0.1108   &  0.0202   &  0.0619  & 0  \\ 
2) $\pi+\sigma(550)+\omega$: ghosts 
        & -0.5894  & -1.9462  & 1.8702   & -0.7717   & -2.3679  & 0  \\  
3) $\pi+\sigma(550)+\omega$: Tree diagrs.
        &  0.9370  & -0.1830  & 0.1977   &  1.0      &   0      & 0  \\      
Sum (A): $1+2+3$
        &  0.3673  & -2.0520  & 2.1790   &  0.2484   & -2.3606  & 0  \\ 
\tableline
4) $\pi+\sigma(770)+\omega$: $\bar{A}_R(\kappa)$ 
        & -0.0061  &  0.0148  & 0.0995   & -0.0202   & -0.0619  & 0  \\
5) $\pi+\sigma(770)+\omega$: ghosts 
        &  0.0358  & -0.0656  & 0.7324   & -0.1197   & -0.3671  & 0  \\
6) $\pi+\sigma(770)+\omega$: Tree diagrs.
        &  0.9679  & -0.1005  & 0.1977   &  1.0      &    0     & 0  \\  
Sum (B): $4+5+6$
        &  0.9976  & -0.1514  & 1.0296   &  0.8602   & -0.4290  & 0  \\ 
\tableline
7) $\pi+\sigma(980)+\omega$: $\bar{A}_R(\kappa)$ 
        & -0.0291  & -0.0424  & 0.0949   & -0.0521   & -0.1600  & 0  \\
8) $\pi+\sigma(980)+\omega$: ghosts 
        &  0.3227  &  0.7724  & 0.3677   &  0.2222   &  0.6818  & 0  \\ 
9) $\pi+\sigma(980)+\omega$: Tree diagrs.
        &  0.9802  & -0.0676  & 0.1977   &  1.0      &    0     & 0  \\
Sum (C): $7+8+9$
        &  1.2737  &  0.6625  & 0.6602   &  1.1701   &  0.5218  & 0  \\ 
\tableline
10) $\pi+\sigma(\infty)+\omega$: $\bar{A}_R(\kappa)$
        & -0.1659  & -0.3935  &  0.0891  & -0.2067   & -0.6342  & 0  \\ 
11) $\pi+\sigma(\infty)+\omega$: ghosts
        &  0.9906  &  2.6735  & -0.0821  &  1.0839   &  0.9314  & 0  \\
12) $\pi+\sigma(\infty)+\omega$: Tree diagrs.
        &  1.0     & -0.01453 &  0.1977  &  1.0      &    0     & 0  \\
Sum (D): $10+11+12$
        &  1.8247  &  2.2654  &  0.2074  &  1.8772   &  2.6917  & 0  \\ 
\tableline
13) NW's work: $\pi+\omega$ - $\bar{A}_R(\kappa)$
        & -0.1659  & -0.3935  &  0.0891  & -0.2067   & -0.6342  & 0  \\ 
14) NW's work: $\pi+\omega$: ghosts 
        &  0.9906  &  2.6735  & -0.0821  &  1.0839   &  3.3259  & 0  \\ 
15) NW's work: $\pi+\omega$: Tree diagrs.
        &    0     & -2.6896  &  0.1977  &    0      & -3.0684  & 0  \\
Sum (E): $13+14+15$: NW's work
        &  0.8247  & -0.4097  &  0.2047  &  0.8772   & -0.3767  & 0 
\end{tabular}
\label{Tab.05}
\end{table}

Tab.~\ref{Tab.05} shows the results for $A^+$ and the scattering lengths 
$a^\pm$ for two cases: $m_\pi=138.03$ MeV and the chiral limit $m_\pi=0$. 
The low-energy theorems impose in the second case that $A^+=1$ 
(in $g^2/M$ units) and $a^\pm=0$. All results are obtained with no form
factors in either the Schwinger-Dyson equation nor in the scattering 
amplitudes. 

The results for the scattering lengths should be compared with the
experimental values given in Eq.~(\ref{Eq.2}), and with the predictions of the
low-energy theorems. The $\pi+\sigma$ system was also studied and the results 
are very similar to the $\pi+\sigma+\omega$ system.
All the contributions for the observable $a^+$ are very 
far from the experimental result; the results of NW are given in Sum (E);
their results for the chiral limit $m_\pi=0$ are correct within 13\% for $A^+$
but they are very large for the $a^+$ observable. Therefore if we examine the
low-energy observables, the ghost poles play no longer a special role.
Our conclusion here is that, as the sum of ghosts plus $\sigma$ contributions 
exceeds by a large amount the experimental values of the $a^+$ scattering 
length, the ghosts are a product of the chosen approximations, which reveal
to be not appropriate at the loop order. 

In order to get rid of the ghosts we change our study to a more 
phenomenological point of view. Using form factors, we investigate the role 
of a ghost free nucleon propagator. One needs first to
recalculate the nucleon pole contribution plus $\sigma$ exchange term using 
one  form factor for each off-shell line at threshold. The sum of all the
contributions depends on two free parameters: $\Lambda_B$ and $\Lambda_m$. 
The last affects the $\sigma$ exchange while the former contributes to the
nucleon pole and to the nucleon self-energy. As mentioned before, the ghosts
disappear for a $\Lambda_B$ less than a critical value $\Lambda_c$.

To fix the values for the cutoff parameters, we first choose a set of 
($\Lambda_m, \Lambda_B$) values such as to reproduce the value $a^+=0$ for 
$m_\pi=0$. Therefore, we get a relation between
the cutoff's. However, this does not uniquely determine their values. To
choose one particular pair of cutoff values, we examined the results for $a^+$
when $m_\pi=138.03$ MeV. It is possible to adjust $\Lambda_m$ and $\Lambda_B$
such that all the different nucleon dressings give the correct chiral limit
$a^+\rightarrow 0$ when $m_\pi=0$; but only with $\Lambda_B$ approximately
1330 MeV one obtains the correct small and negative result for $a^+$ when
$m_\pi=138.03$ MeV.
Tab.~\ref{Tab.06} shows the values for $\Lambda_m$ for $\pi+\sigma$ and
$\pi+\sigma+\omega$ dressings and different $\sigma$ masses. There is one
delicate point here, namely the value for $\Lambda_m$ when
$m_\sigma\rightarrow\infty$. We handle this by making these two quantities go 
to infinity at the same time, keeping the ratio between them constant to
preserve the chiral limit $a^+=0$ when $m_\pi=0$. Our main result here is that
$m_\sigma/\Lambda_m\approx 0.67$ in order to reproduce the results for $a^+$
at threshold and at the chiral limit. In doing this, we are constructing a 
phenomenological model which respects the low-energy theorems in $\pi N$
scattering. It is clear also that we can not make any statement about chiral 
symmetry, since the phenomenological form factors violate the chiral
Ward-Takahashi identities.

\begin{table}
\caption{Values of $\Lambda_m$ adjusted to reproduce the scattering length
$a^+$ for for $\pi+\sigma$ and $\pi+\sigma+\omega$ dressings at different 
$\sigma$ masses.}
\begin{tabular}{lccrr}
Process               &\multicolumn{2}{c}{$\pi+\sigma$ system}    
                      &\multicolumn{2}{c}{$\pi+\sigma+\omega$ system} \\ 
\cline{2-3} 
\cline{4-5}
$\sigma$ masses (MeV) & $\Lambda_m$ & $m_\sigma / \Lambda_m$ 
                      & $\Lambda_m$ & $m_\sigma / \Lambda_m$  \\ 
\tableline
$m_\sigma=550$  &  822.2825  & 0.66887    &   821.4860  & 0.66952  \\
$m_\sigma=770$  & 1145.079   & 0.67244    &  1143.978   & 0.67309  \\
$m_\sigma=980$  & 1455.984   & 0.67308    &  1454.585   & 0.67373  \\
$m_\sigma\rightarrow\infty$ 
                &   -        & 0.67338    &     -       & 0.67403
\end{tabular}
\label{Tab.06}
\end{table}

\begin{table}
\caption{Same as in TABLE IV, but using form factors.} 
\vspace{0.3cm}
\begin{tabular}{lrrrrrr} 
Contribution    & $A^+$ & $m_\pi\cdot B^+$& $A^-$ & $m_\pi\cdot B^-$& $a^+$ 
                  & $a^-$  \\ 
                & $g^2/M$ & $g^2/M$ & $g^2/M$ & $g^2/M$ & (fm)     & (fm)   \\ 
\tableline
1) $\pi+\sigma(550)+\omega$: $\bar{A}_R(\kappa)$
                & 0.3744  &  0.0249 & 0.1335  & 0.0644  &  1.06820 & 0.5292 \\
2) $\pi+\sigma(550)+\omega$: Tree diagrs.
                & 0.3986  & -0.8335 &   0     & 0.0613  & -1.16336 & 0.1639 \\
Sum (A): $1+2$
                & 0.7730  & -0.8084 & 0.1335  & 0.1267  & -0.09516 & 0.6931 \\ 
\tableline 
3) $\pi+\sigma(770)+\omega$: $\bar{A}_R(\kappa)$
                & 0.3811  &  0.0249 & 0.1329  & 0.0605  &  1.08513 & 0.5173 \\
4) $\pi+\sigma(770)+\omega$: Tree diagrs.
                & 0.4084  & -0.8335 &    0    & 0.0613  & -1.13713 & 0.1639 \\
Sum (B): $3+4$  
                & 0.7895  & -0.8089 & 0.1329  & 0.1218  & -0.05200 & 0.6812 \\
\tableline
5) $\pi+\sigma(980)+\omega$: $\bar{A}_R(\kappa)$
                & 0.3821  &  0.0246 & 0.1327  & 0.06001 &  1.08771 & 0.5155 \\
6) $\pi+\sigma(980)+\omega$: Tree diagrs.
                & 0.4130  & -0.8335 &   0     & 0.0613  & -1.12486 & 0.1639 \\
Sum (C):  $5+6$
                & 0.7951  & -0.8089 & 0.1327  & 0.1213  & -0.03715 & 0.6794 \\
\tableline 
7) $\pi+\sigma(\infty)+\omega$: $\bar{A}_R(\kappa)$
                & 0.3815  &  0.0246 & 0.1329  & 0.0601  &  1.08640 & 0.5163 \\
8) $\pi+\sigma(\infty)+\omega$: Tree diagrs.
                & 0.4211  & -0.8335 &    0    & 0.0613  & -1.10326 & 0.1639 \\
Sum (D): $7+8$
                & 0.8026  & -0.8089 & 0.1329  & 0.1214  & -0.01686 & 0.6802 
\end{tabular}
\label{Tab.08}
\end{table}

Tab.~\ref{Tab.08} presents the results for the observables for $m_\pi=138.03$
MeV. All cases studied have the correct chiral limit $A^+=g^2/M$ and $a^\pm=0$
for $m_\pi=0$. The results for the case $\pi+\sigma$ are not shown since they
are similar to the ones for the $\pi+\omega+\sigma$ case.

Observable $A^+$ receives contributions mainly from (a) the
$\sigma$-exchange, Fig.~\ref{Fig.6}(c), which depends directly on $\Lambda_m$,
and (b) from the nucleon self-energy $\bar A_R(\kappa)$, which is weakly
dependent on $\Lambda_B$ and is almost independent from the $\sigma$ mass. 
The results show that $\sigma$ exchange and the self-energy contribution
$\bar A_R(\kappa)$ contribute approximately 50\% each. 

The results for $a^+$
depend on the sum of $A^+$ and $B^+$. Observable $B^+$ receives contributions
from the nucleon Born part, Fig.~\ref{Fig.6}(a),(b), and from the spectral
function $\bar A_R(\kappa)$. The contribution from the spectral function is
almost constant and very small. The contribution from nucleon pole term 
depends very weakly on $\Lambda_B$. 

Observable $a^-$ is not well adjusted due to the huge contribution from
$\bar{A}_R(\kappa)$, being almost 75\% of the final result. This huge 
contribution comes from negative $\kappa$ part of the spectral function. 
We checked this point by doing $\bar{A}_R(\kappa)=0$ for $\kappa<0$ by hand and
get a result 5 times smaller for $a^-$.
As discussed previously, the negative $\kappa$ enhancement is an effect
due to the form factors. This is a limitation of the model,
since as we fix the cutoffs to reproduce the isoscalar low-energy observables, 
we can not reproduce the isovector ones.

\subsection{Partial waves and Phase-shifts}

The total amplitude for $\pi N$ scattering may be decomposed into the 
isospin $\frac{3}{2}$ and $\frac{1}{2}$ channels. The isospin $\frac{3}{2}$
and $\frac{1}{2}$ amplitudes are related to the symmetric and 
antisymmetric amplitudes by

\begin{eqnarray}
O^{(3/2)}&=&O^{(+)}-O^{(-)}\;, \nonumber\\
O^{(1/2)}&=&O^{(+)}+2O^{(-)}\;.
\label{op}
\end{eqnarray}

\noindent
where $O$ can be $A$ or $B$. Therefore, 
the $T$ matrix can be decomposed into good isospin and total angular momentum
channels to reveal the existence of any resonances. The angular momentum 
decomposition of $T$ leads to~\cite{Rom69}:
\begin{eqnarray}
f_{\ell\pm}^I(W)=\frac{1}{16\pi W}&&\int_{-1}^{1}dx\left\{(E+M)\left[A^I(W,x)+
(W-M)B^I(W,x)\right]P_\ell(x)\right.\nonumber \\
&&\left.+(E-M)\left[-A^I(W,x)+(W+M)B^I(W,x)\right]P_{\ell\pm 1}(x)\right\},
\label{partial}
\end{eqnarray}
where $f$ is the partial wave amplitude, $\ell$ is the orbital angular momentum,
and $I$ is the isospin.

From Eq.~(\ref{op}) one sees that the $I=\frac{1}{2}$ channel
is the result of a sum over symmetric and antisymmetric  channels, which
led to an imaginary part on the partial amplitude due to the pole for
$\kappa=\pm\sqrt{s}$ on $A^\pm$ and $B^\pm$ amplitudes, 
Eqs.~(\ref{A1},\ref{A3}). This fact does not happen in the $I=\frac{3}{2}$
channel due to the amplitude subtraction in Eq.~(\ref{op}), which cancels the
$s-\kappa^2$ denominator. Thus, this approach cannot represent any of the
$I=\frac{3}{2}$ resonances, unless some kind of unitarization is made.

We can now define the scattering length ($a$) in terms of the partial wave 
amplitude ($f$). For each $\ell$ value we can expand $f$ in terms of $q$
near the threshold ($q=0$) and write
\begin{equation}
a_{\ell\pm}^I=\lim_{q\rightarrow 0}\frac{f_{\ell\pm}^I}{q^{2\ell}}.
\label{comp}
\end{equation} 
One observes that for the $S$ wave ($\ell=0$), the scattering length is 
the partial wave amplitude at the threshold, $a_{0+}^I=f_{0+}^I(q=0)$.

From the unitarity condition on the $S$ matrix, we can relate the partial wave
amplitude ($f$) and the phase shifts ($\delta$) as

\twocolumn

\begin{equation}
f_{\ell\pm}^I=\frac{e^{\imath\delta_{\ell\pm}^I}}{q}\sin\delta_{\ell\pm}^I\;.
\label{delta}
\end{equation}

The optical theorem identifies the total cross section with the imaginary part
of the scattering amplitude. We unitarize the amplitude following the method
of Olsson and Osypowski~\cite{OO75}. In their method, the real amplitude
that arises from the model, $m$, is associated with the real part of the
partial wave amplitude $f$ as:

\begin{eqnarray}
m_{\ell\pm}^I&=&\mbox{ Re }f_{\ell\pm}^I=\frac{\sin\,2\delta_{\ell\pm}^I}{2q}
\\
\delta_{\ell\pm}^I&=&\frac{1}{2}\mbox{ arc sin}\left(2\,q\,m_{\ell\pm}^I\right)
\label{defa}
\end{eqnarray}

\begin{figure}
\vspace{0.5cm}
\epsfxsize=7.2cm
\epsfysize=9cm
\centerline{\epsffile{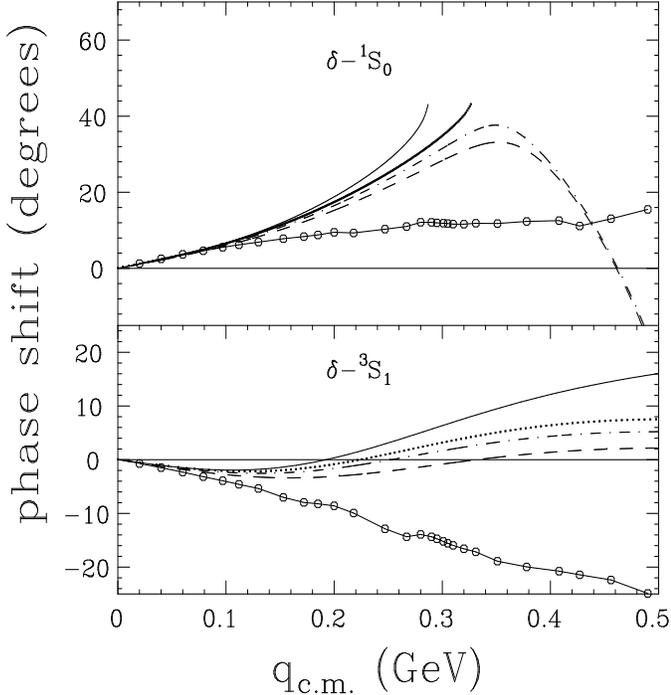}}
\vspace{1cm}
\caption{Phase shifts for $\ell=0$ waves. The curves represent the choices
for the $\sigma$-meson mass: 550 MeV (dashed), 770 MeV (dot-dashed), 980 MeV
(dotted), and $m_\sigma\rightarrow\infty$ (solid curve). Data are from
\protect\cite{Hol83}.}
\label{Fig.16}
\end{figure}

Fig.~\ref{Fig.16} presents the results for the  phase shifts for $\ell=0$ waves.
The agreement near threshold is very good, and the model fails for 
higher energies. The other phase-shifts show the same trend: they are well 
described near threshold, but as energy increases the agreement with data
becomes poor.

A much better agreement for the phase-shifts can be achieved if we choose 
another set of values for the cutoffs~\cite{Priv}, but the good agreement 
for the low-energy observables is destroyed. 
 
\section{CONCLUSIONS}
\label{sec:conclusions}

Previous studies of $\pi N$ scattering by Nutt and Wilets showed that in a
model with $\pi$ and $\omega$ mesons the inclusion of the ghost 
poles lead to a much better agreement with experiment than if the poles were
neglected. The fact that the Adler consistency condition could be
satisfied in this approach is very intriguing since while one would
expect that low energy observables should be insensitive to the ultraviolet
(after renormalization), ghosts are a consequence of the ultraviolet behavior 
of the interaction kernel in the Hartree-Fock approximation. However, 
when we combine the $A^+$ and the $a^+$ observables, the ghost poles play no 
longer a special role, since the $a^+$ experimental result can not be 
reproduced by this first approach. One natural solution to this problem is to
modify the model in order to include chiral symmetry. 
In this paper we apply this idea in the light of the linear $\sigma$ 
model augmented with the $\omega$ meson. The solution of the nucleon 
Schwinger-Dyson equation obtained in the Hartree-Fock approximation in 
this approach also contains a pair of ghost poles.
Moreover, the sum of ghosts plus $\sigma$ contribution 
exceeds by a large amount the experimental values of the $a^+$ scattering 
length, leading to the conclusion that the ghosts are due to the 
approximations adopted. By definition, the scattering amplitude for the 
$\pi N$ interaction in the linear $\sigma$ model has the correct chiral limit
at the three-level calculations. As our calculations include loops, one needs
to add additional terms in the lagrangian, to make sure that the resulting 
amplitude has the correct chiral limit. At the same time, it is possible 
to choose the constants of these additional terms in order to kill the 
complex poles. We intend to present this study in a near future.

In order to eliminate the ghosts, we used phenomenological form factors 
at the vertex interactions. In softening the ultraviolet
by means of form factors it is possible to obtain qualitative agreement with 
experimental data of observables at low energies. The first observation  
from our study is that the spectral function $\bar A(\kappa)$ for 
negative $\kappa$ is strongly enhanced by the form factors and this affects 
some of the observables. In particular, the negative $\kappa$ enhancement 
increases the isospin antisymmetric scattering length $a^-$. Another lesson
from this phenomenological model is that the phase shifts are well described
only at threshold if we keep the cutoff values that reproduces the low-energy
observables. One first conclusion is that it is almost impossible to 
reproduce together the observables at low and intermediate energies  
in $\pi N$ interaction 
with this simple phenomenological model, even including the fluctuations in 
the nucleon propagator. 

In view of the compelling evidences that chiral symmetry is 
a fundamental symmetry of the strong interactions, the problem of the 
appearance of ghost poles at low energies in commonly used truncation schemes 
in field theoretic models has to be very carefully examined before definite 
conclusions can be drawn on the validity of a particular model used for the 
description of the data. One has believed that Chiral Perturbation Theory 
(CHPT) is 
the unique solution, in the near future, to fill these requirements, since it
clearly states how to add loops and keep the chiral limits of the low-energy 
observables. However, to describe data at intermediate energies, one needs
to evaluate diagrams with 2 or maybe more loops, which are very 
involved due to the regularization process done in every order. Moreover, 
the number of unknown constants increase with the number of loops, 
compromising the predict power of the theory.

We believe that much still remains to be studied with respect to
the interplay of chiral symmetry and the ultraviolet behavior of the model
interactions. One particular issue is the role of vertex corrections in
the Schwinger-Dyson equation and the requirement that these should satisfy
the chiral Ward-Takahashi identities. Such constraints are expected to be
relevant at the low-energy region of the kernel interactions and as such
can be of importance for a better fit to experimental data at low energies.

\acknowledgements

We thank Prof. M. Robilotta and Prof. T. Cohen for  helpful conversations 
about $\pi N$ scattering and chiral symmetry. This work was partially 
supported by U.S. Department of Energy. The work of C.A. da Rocha, 
was supported by CNPq Grant No. 200154/95-8 and (in the early stages of this 
work) by FAPESP Grant No. 92/5095-1, both Brazilian agencies.

\end{document}